\documentclass[prl,twocolumn,showpacs]{revtex4}

\usepackage{graphicx}           
\usepackage{dcolumn}            
\usepackage{bm}                 

\begin{document}


\title{Polarization Properties of Single Quantum Dots in Nanowires}

\author{M.~H.~M.~van~Weert}
\author{N.~Akopian}
\affiliation{Kavli Institute of Nanoscience Delft, Delft University
of Technology, Delft 2628CJ, The Netherlands}
\author{F.~Kelkensberg}
\altaffiliation{Current address: FOM-Institute AMOLF, Amsterdam, The
Netherlands}
\author{U.~Perinetti}
\author{M.~P.~van~Kouwen}
\affiliation{Kavli Institute of Nanoscience Delft, Delft University
of Technology, Delft 2628CJ, The Netherlands}
\author{J.~G\'omez Rivas}
\affiliation{FOM Institute for Atomic and Molecular Physics AMOLF,
c/o Philips Research Laboratories, 5656AE, The Netherlands}
\author{M.~T.~Borgstr\"om}
\altaffiliation{Current address: Lund University, Lund, Sweden}
\author{R.~E.~Algra}
\altaffiliation{Also at: Materials Innovation Insitute, Delft, The
Netherlands, and Solid State Chemistry, Radboud University Nijmegen,
Nijmegen, The Netherlands}
\author{M.~A.~Verheijen}
\author{E.~P.~A.~M.~Bakkers}
\affiliation{Philips Research Laboratories, Eindhoven 5656AE, The
Netherlands}
\author{L.~P.~Kouwenhoven}
\author{V.~Zwiller}
\email{v.zwiller@tudelft.nl}
\affiliation{Kavli Institute of
Nanoscience Delft, Delft University of Technology, Delft 2628CJ, The
Netherlands}


\begin{abstract}

We study the absorption and emission polarization of single
semiconductor quantum dots in semiconductor nanowires. We show that
the polarization of light absorbed or emitted by a nanowire quantum
dot strongly depends on the orientation of the nanowire with respect
to the directions along which light is incident or emitted. Light is
preferentially linearly polarized when directed perpendicular to the
nanowire elongation. In contrast, the degree of linear polarization
is low for light directed along the nanowire. This result is vital
for photonic applications based on intrinsic properties of quantum
dots, such as generation of entangled photons. As an example, we
demonstrate optical access to the spin states of a single nanowire
quantum dot.

\end{abstract}

 \maketitle

Semiconductor quantum dots (QDs) are sources of
single~\cite{imamoglu_science_2000,yamamoto_prl_2001} and entangled
photons~\cite{petroff_prl_2006,shields_njp_2006,schmidt_njp_2007},
single electrons~\cite{zrenner_nature_2002}, and are naturally
integrated with modern semiconductor electronics. Incorporating them
in semiconducting nanowires
(NWs)~\cite{imamoglu_nanolett_2005,bakkers_nanolett_2007,harmand_nanolett_2007,fukui_nanolett_2007,gayral_nanolett_2008}
brings additional unique features such as natural alignment of
vertically stacked QDs~\cite{petroff_science_1999} and an inherent
one-dimensional channel for charge carriers. Furthermore, the
unprecedented material and design freedom makes them very attractive
for novel opto-electronic
devices~\cite{lieber_science_2001,yang_science_2001,lieber_nature_2001_inpnws,lieber_nature_2003,samuelson_nanolett_2006,capasso_nanolett_2006}
and quantum information science in general~\cite{loss_pra_1998,
samuelson_nanolett_2004_fewelectron,lieber_nanolett_2005,ensslin_apl_2006}.
However, access to intrinsic spin and polarization properties of a
QD in a NW has never been demonstrated, most likely due to an
insufficient quality of the NW QDs. Moreover, the NW geometry
strongly affects the polarization of photons emitted or absorbed by
a NW QD, and thus becomes the main obstacle for applications based
on intrinsic spin or polarization properties of QDs such as an
electron spin memory~\cite{finley_nature_2004} or generation of
entangled photons~\cite{petroff_prl_2006}. Indeed, it has been shown
that luminescence of pure NWs is highly linearly polarized and the
polarization direction is parallel to the NW
elongation~\cite{lieber_science_2001,samuelson_nanolett_2006,rivas_optlett_2007}.

In this Letter we demonstrate that by directing light along the NW
elongation we can access intrinsic spin and polarization properties
of a QD in a NW. We introduce a theoretical model which intuitively
explains our experimental findings and shows how polarization is
affected by various parameters such as NW diameter, dielectric
constant of the surroundings and photon wavelength. As an example,
we demonstrate access to the spin properties of a Zeeman split
exciton in a QD by measuring the right- and left-hand circular
photon polarization.

\begin{figure}
\begin{center}
\includegraphics[width=0.9\linewidth]{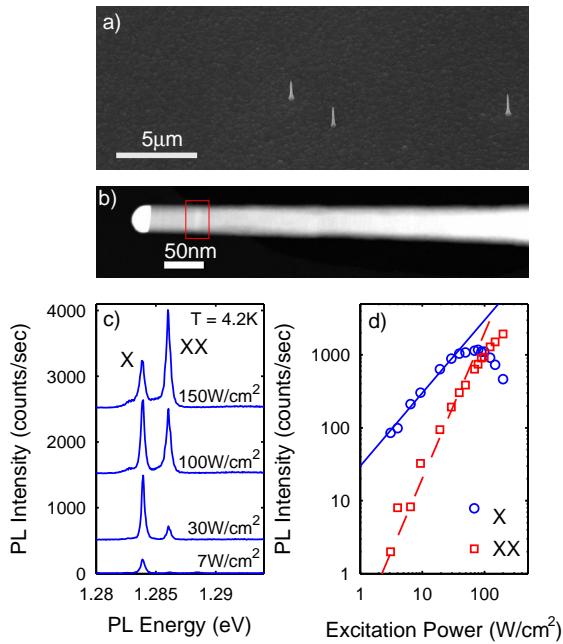}
\caption{\label{fig:fig1}(color online) a) SEM image of an as-grown
sample with low density  of standing NW QDs. b) TEM image of a NW
QD. The QD is indicated by the red rectangle. The low contrast of
the QD is due to the low As concentration. c) PL spectra of a single
standing NW QD for various excitation powers. The two emission lines
denoted with X and XX correspond to the exciton and biexciton
emission, respectively. d) PL intensity of X and XX transitions as a
function of the excitation power. The solid (dashed) line is a guide
to the eye for linear (quadratic) power dependence.}
\end{center}
\end{figure}

For our experiments we used single InAs$_{0.25}$P$_{0.75}$ QDs
embedded in InP NWs, grown by means of metal-organic vapor-phase
epitaxy~\cite{hiruma_apl_1991,lieber_nature_2002,bakkers_jacs_2006,bakkers_nnano_2007}.
Colloidal gold particles of 20~nm diameter were deposited on a
(111)B InP substrate as catalysts for vertical NW growth. The
diameter of the NW and the QD was controlled by the gold particle
size, while the NW density was set by the gold particle density on
the substrate. The QD height and NW length were determined by growth
time~\cite{bakkers_nanolett_2007}. By controlling diameter, height,
and As concentration one can tune the QD emission in the wide range
of 900~nm to 1.5~$\mu$m~\cite{harmand_nanolett_2007}. Under
appropriate growth conditions we were able to grow a sample with low
density of NWs with a single QD that emits around 950 nm. In
Fig.~\ref{fig:fig1}a) we show a scanning electron microscope (SEM)
image of an as-grown sample with a low density of NWs, enabling
optical study of a single NW QD. The NW length is about 2~$\mu$m. A
transmission electron microscope (TEM) image of a NW QD is shown in
Fig.~1b). The QD is typically 9~nm high with a diameter of 31~nm,
determined by energy dispersive X-ray analysis in a TEM. To simplify
TEM studies, the InP section following QD growth was reduced, while
for the samples used in our experiments, the QD was centered in the
NW.

As-grown samples with vertically oriented NWs, referred as standing
NWs, were used for experiments where the excitation (emission) was
directed (measured) along the NW. We also transferred NWs to a Si
substrate with a 290 nm SiO$_2$ top layer, referred as lying NWs,
for the experiments where light was directed perpendicular to the
NW. Both samples were studied under the same experimental
conditions. Micro photoluminescence (PL) studies were performed at
4.2~K. The NW QDs were excited with a linearly polarized 532~nm cw
laser focused to a spot size of 1~$\mu$m using a microscope
objective with a numerical aperture NA=0.85. The PL signal was
collected by the same objective and was sent to a spectrometer,
which dispersed the PL onto a nitrogen-cooled silicon array
detector, enabling 50 $\mu$eV resolution. Linear and circular
polarizations were determined with a fixed polarizer together with a
half- and quarter-waveplate, respectively.

In Fig.~\ref{fig:fig1}c) we show PL spectra of a single standing NW
QD for various excitation powers. At low excitation power only one
emission line is visible at 1.284~eV, denoted as X. With increasing
excitation power a second emission line, denoted as XX, emerges
2.2~meV above X. The PL intensities of X and XX as a function of
excitation power, represented in Fig.~\ref{fig:fig1}d), show that X
(XX) increases linearly (quadratically) with excitation power and
saturates at high excitation powers. This behavior is typical for
the exciton and biexciton under cw excitation. The spectral
linewidths for various NW QDs in our samples are in the range of
50-200 $\mu$eV, demonstrating good quality of the NW QDs. Similar
spectral characteristics are also found for lying NW QDs.

\begin{figure}
\begin{center}
\includegraphics[width=0.9\linewidth]{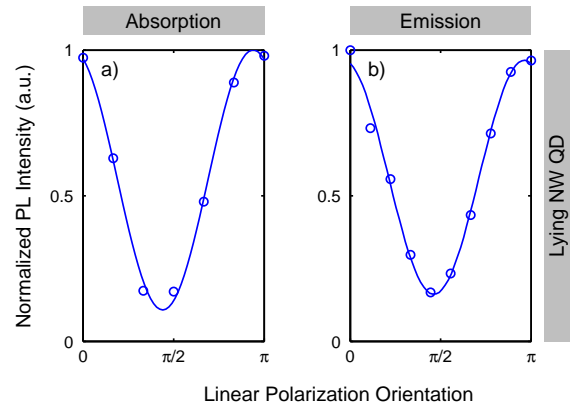}
\caption{\label{fig:fig2}(color online) Integrated PL intensity of
the exciton in a lying NW QD a) as a function of the orientation of
the linear excitation polarization and b) by probing the orientation
of the linear emission polarization. Zero angle of orientation was
chosen to match the NW elongation, determined by imaging. The solid
curves are $\sin^2$ fits used to extract the degree of linear
polarization from minima and maxima of the fit.}
\end{center}
\end{figure}

In Fig.~\ref{fig:fig2} we show the integrated PL intensity of the
exciton in a lying NW QD as a function of linear excitation and
emission polarization orientation. In Fig.~\ref{fig:fig2}b) the
excitation polarization is set parallel to the NW elongation.
However, the emission polarization is independent of the excitation
polarization. We fit this data to a $\sin^2$-function with
amplitude, offset, and phase as fit parameters. The degree of linear
polarization, \mbox{$\rho =
\frac{I_{max}-I_{min}}{I_{max}+I_{min}}$}, is obtained from the
maxima and minima of the fit. Clearly, lying NW QDs show a large
degree of linear polarization parallel to the NW, for both
absorption and emission.


\begin{figure}
\begin{center}
\includegraphics[width=0.9\linewidth]{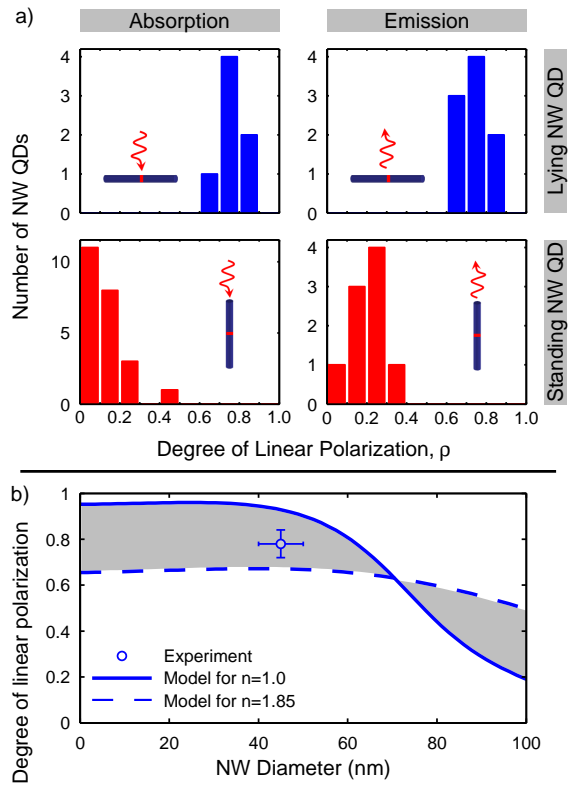}
\caption{\label{fig:fig3}(color online) a) Statistics of the degree
of linear polarization for lying and standing NW QDs. The lying NW
QDs (shown in blue) show high degree of linear polarization, about
0.8 and 0.7, for absorbed and emitted light, respectively. In
contrast, the degree of linear polarization is low for standing NW
QDs (shown in red), about 0.1 and 0.2 for absorbed and emitted
light, respectively. Insets illustrate the NW orientations with
respect to the light direction. b) Calculated degree of linear
polarization in absorption as a function of NW diameter. In this
case incident light is directed perpendicular to the NW elongation,
as for the lying NW QDs. For the solid (dashed) curve an effective
refractive index of $n=1$ ($n=1.85$) is used.}
\end{center}
\end{figure}

Polarization-sensitive PL measurements were performed on a number of
NW QDs, both lying and standing, which are shown in
Fig.~\ref{fig:fig3}a). The upper two graphs show the statistics of
absorption and emission for lying NW QDs, for which absorbed
(emitted) light is directed (measured) perpendicular to the NW
elongation. The average degree of linear polarization is
\mbox{$\rho_{abs}=0.78 \pm 0.06$} \mbox{($\rho_{emi}=0.72 \pm
0.08$)} for absorbed (emitted) light.

The same measurements were performed on standing NW QDs, for which
light absorbed (emitted) by the QD is directed (measured)
parallel to the NW elongation. The results are shown in the lower
two graphs of Fig.~\ref{fig:fig3}a) and give a degree of linear
polarization for absorbed (emitted) light of $\rho_{abs}=0.1$
($\rho_{emi}=0.2$). This low value is expected for the symmetric
structure of the NWs for light incident or emitted in the direction
along the NW elongation. In addition, the measured degree of
circular polarization of the QD emission is also low, 0.1.



To explain our experimental results for absorption by a lying NW QD,
we use Mie theory for light scattering on dielectrics of cylindrical
shape~\cite{bohren,ruda_jap_2006}. The NW is modeled as an infinite
cylinder. This approximation is valid as long as the nanowire
diameter is much smaller than its length, as in our case. We
calculate the scattering and absorption of light using the scalar
wave equation in cylindrical coordinates. We consider two cases:
incident electric field parallel or perpendicular to the NW
elongation. We extract the scattered field for both polarizations
from the wave equation, using the boundary conditions at the
interface of two different dielectrics~\cite{jackson}, i.e., at the
NW surface. From this field, we calculate for both cases the
cross-sections $Q_{sca}$ and $Q_{ext}$ for scattering and
extinction. The absorption cross section is obtained by $Q_{abs} =
Q_{ext}-Q_{sca}$ for both polarizations. To take into account the
illumination of the NW through a microscope objective, we integrate
the average absorption cross section over angles of incidence with
respect to the NW elongation comprised between $\arccos$(NA) and
$\pi/2$.

Mie theory assumes the surrounding of the NW as a homogeneous
medium, which differs from our situation where the NW is lying on a
substrate. Therefore, to approximate the effect of the substrate we
consider the nanowire as being embedded in a medium with an
effective refractive index, i.e., an average of the refractive
indices of the different media surrounding the NW: vacuum, SiO$_2$,
and Si. The outcome of the calculations, assuming an effective
refractive index of \mbox{$n_{eff} = 1.85 = 0.5n_{vacuum} +
0.25n_{SiO_2} + 0.25n_{Si}$} is represented by the dashed curve in
Fig.~\ref{fig:fig3}b). As an upper limit we consider the NW in
vacuum, thus ignoring the substrate, which is represented by the
solid curve in Fig.~\ref{fig:fig3}b). For the calculations we use
our actual excitation wavelength of 532~nm. Our experimental value
lies in between the two curves, having a surrounding refractive
index consisting of a mixture of vacuum, Si, and SiO$_2$. As can be
seen in Fig.~\ref{fig:fig3}b) one can increase the degree of linear
polarization by measuring NW QD in vacuum, or decrease it by
increasing the NW diameter. However, in the latter case the
advantage of the one-dimensional channel of the device is reduced as
well.

The NW geometry is not the only source of polarization anisotropy.
Calculations by Niquet and Mojica~\cite{niquet_prb_2008} show that
the polarization properties are strongly affected by the aspect
ratio of the QD dimensions, due to strain originating from the
lattice mismatch between the NW and the QD. However, in our case the
strain is negligible due to the low phosphorus content and the main
contribution to polarization anisotropy comes from the NW geometry.

%

\begin{figure}
\begin{center}
\includegraphics[width=0.9\linewidth]{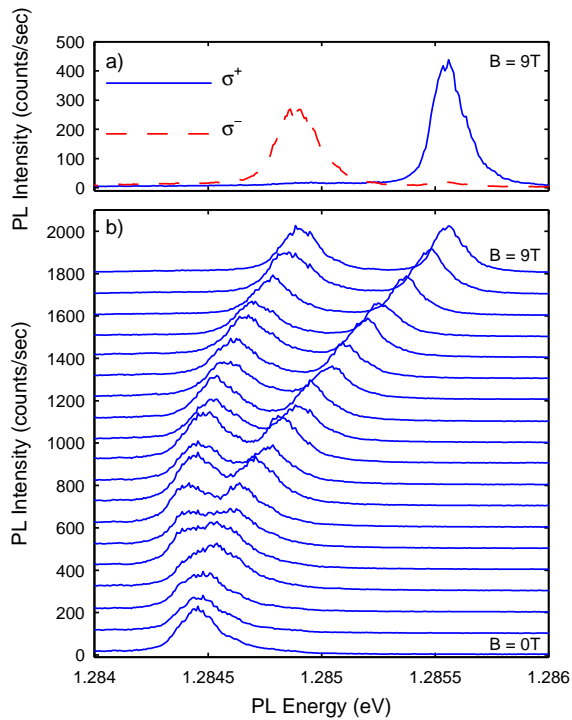}
\caption{\label{fig:fig4}(color online) a) Polarization sensitive PL
of a standing NW QD at 9 T. The solid (dashed) curve represent
right- (left-) hand circularly polarized exciton emission, denoted
by $\sigma^{+}$ ($\sigma^{-}$). b) PL of a standing NW QD under
external magnetic field. Magnetic field is varied between 0 and 9 T
in steps of 0.5 T.}
\end{center}
\end{figure}

The low degree of linear polarization for vertical NW QDs implies
that the NW does not affect the polarization of photons radiated
from the QD, thus allowing access to intrinsic spin and polarization
properties of NW QDs. To experimentally demonstrate this we measure
the polarization of photons emitted from a Zeeman split exciton in a
NW QD. In Fig.~\ref{fig:fig4}b) we show PL of the NW QD exciton
under external magnetic field $B$ in the Faraday configuration ($B
\parallel$~NW). Clearly, the exciton undergoes a Zeeman splitting and
diamagnetic shift. From these measurements we extract the exciton
$g$-factor, $g = 1.3 \pm 0.1$, which is typical for our sample.
Polarization properties of the split exciton lines at 9~Tesla are
shown in Fig.~\ref{fig:fig4}a). The high energy emission line is
right-hand circularly polarized, while the lower energy emission
line is left-hand circularly polarized. These two polarization
states, $\sigma^+$ and $\sigma^-$, correspond to the two different
spin states of the exciton $\downarrow \Uparrow$ and $\uparrow
\Downarrow$, respectively, where $\uparrow (\Uparrow)$ represents a
spin-up electron (hole) and $\downarrow (\Downarrow)$ represents a
spin-down electron (hole). These results show that the emission
polarization of a QD is not obscured by the NW geometry, and that
electron-hole spin states in a QD are properly converted to the
polarization states of the emitted photons. In contrast,
measurements on a lying NW QD show that the two emission lines of a
Zeeman split exciton are strongly linearly polarized, with respect
to the NW elongation, as expected (not shown).

To conclude, we have correlated the polarization of light absorbed
and emitted by a QD embedded in a NW with its propagation direction
with respect to the NW elongation.
We found that the polarization of the absorbed (emitted) light, when
directed (measured) perpendicular to the NW elongation, is affected
by the NW geometry and is strongly linearly polarized along the NW.
In contrast, this is not the case for the configuration where the
absorbed (emitted) light is directed (measured) parallel to the NW
elongation. In this configuration the intrinsic spin and
polarization of the NW QD can be accessed. We have demonstrated this
by measuring polarization states of photons radiated from a Zeeman
split exciton in a NW QD.

\begin{acknowledgments}
We acknowledge W.~G.~G.~Immink for technical assistance and
F.~Holthuysen for providing the SEM image. This work was supported
by the European FP6 NODE (015783) project. The work of JGR was
supported by the Netherlands Foundation Fundamenteel Onderzoek der
Materie (FOM) and the Nederlandse Organisatie voor Wetenschappelijk
Onderzoek (NWO) and is part of an industrial partnership program
between Philips and FOM. The work of REA was carried out under
project number MC3.0524 in the framework of the strategic research
program of the Materials Innovation Institute (M2I)
\mbox{(www.m2i.nl)}.
\end{acknowledgments}


\end{document}